\documentclass[prd,preprint,aps,showpacs]{revtex4}
\usepackage{graphics}
\usepackage{amsfonts}
\usepackage{placeins}
\usepackage{lscape}
\usepackage{rotating}
\usepackage{float}
\newcommand{\newc}{\newcommand}
\newc{\gsim}{\lower.7ex\hbox{$\;\stackrel{\textstyle>}{\sim}\;$}}
\newc{\lsim}{\lower.7ex\hbox{$\;\stackrel{\textstyle<}{\sim}\;$}}
\newc{\gev}{\,{\rm GeV}}
\newc{\mev}{\,{\rm MeV}}
\newc{\ev}{\,{\rm eV}}
\newc{\kev}{\,{\rm keV}}
\newc{\tev}{\,{\rm TeV}}
\newc{\mz}{m_Z}
\newc{\mpl}{M_{Pl}}
\newc{\chifc}{\chi_{{}_{\!F\!C}}}
\newc\order{{\cal O}}
\newc\CO{\order}
\newc\CL{{\cal L}}
\newc\CY{{\cal Y}}
\newc\CH{{\cal H}}
\newc\CM{{\cal M}}
\newc\CF{{\cal F}}
\newc\CD{{\cal D}}
\newc\CN{{\cal N}}
\newc{\eps}{\epsilon}
\newc{\re}{\mbox{Re}\,}
\newc{\im}{\mbox{Im}\,}
\newc{\invpb}{\,\mbox{pb}^{-1}}
\newc{\invfb}{\,\mbox{fb}^{-1}}
\newc{\yddiag}{{\bf D}}
\newc{\yddiagd}{{\bf D^\dagger}}
\newc{\yudiag}{{\bf U}}
\newc{\yudiagd}{{\bf U^\dagger}}
\newc{\yd}{{\bf Y_D}}
\newc{\ydd}{{\bf Y_D^\dagger}}
\newc{\yu}{{\bf Y_U}}
\newc{\yud}{{\bf Y_U^\dagger}}
\newc{\ckm}{{\bf V}}
\newc{\ckmd}{{\bf V^\dagger}}
\newc{\ckmz}{{\bf V^0}}
\newc{\ckmzd}{{\bf V^{0\dagger}}}
\newc{\X}{{\bf X}}
\newc{\bbbar}{B^0-\bar B^0}

\newc{\sgn}{\mbox{sgn}\,}
\newc{\m}{{\bf m}}
\newc{\msusy}{M_{\rm SUSY}}
\newc{\munif}{M_{\rm unif}}
\newc{\slepton}{{\tilde\ell}}
\newc{\Slepton}{{\tilde L}}
\newc{\sneutrino}{{\tilde\nu}}
\newc{\selectron}{{\tilde e}}
\newc{\stau}{{\tilde\tau}}
\def\beq{\begin{equation}}
\def\eeq{\end{equation}}
\def\bea{\begin{eqnarray}}
\def\eea{\end{eqnarray}}
\newc{\ie}{{\it i.e.}}          \newc{\etal}{{\it et al.}}
\newc{\eg}{{\it e.g.}}          \newc{\etc}{{\it etc.}}
\newc{\cf}{{\it c.f.}}
\def\beqn{\begin{eqnarray}}
\def\eeqn{\end{eqnarray}}
\relax
\newcommand{\ba}[1]{\begin{array}{#1}}
\def\ea{\end{arroy}}

\def\beq{\begin{equation}}
\def\eeq{\end{equation}}
\def\bea{\begin{array}}
\def\eea{\end{array}}

\def\ifb{ {\rm fb}^{-1} }

\def\to{\rightarrow}

\def\[{\left[}
\def\]{\right]}
\def\({\left(}
\def\){\right)}

\def\U1em{{U(1)_{\rm em}}}
\def\to{\rightarrow}

\def\CL{{\cal C}_L}

\def\sq2{\sqrt{2}}

\def\End{\end{document}}

\def\slash#1{\rlap{$#1$}/} 
\def\Dsl{\,\raise.65ex\hbox{/}\mkern-03.5mu D} 
\def\delsl{\raise.15ex\hbox{/}\kern-.57em\partial}
\def\Ksl{\hbox{/\kern-.6700em\rm K}}
\def\Asl{\hbox{/\kern-.1500em \rm A}}
\def\Qsl{\hbox{/\kern-.6000em\rm Q}}
\def\gradsl{\hbox{/\kern-.6501em$\nabla$}}
\def\bar#1{\overline{#1}}

\begin{document}
\draft
\title{Electroweak right-handed neutrinos and new Higgs signals at the LHC}
\author{J.L. D\'{\i}az-Cruz$^{(a)}$, O. F\'elix-Beltr\'an$^{(b)}$, A. Rosado$^{(c)}$ and S. Rosado-Navarro$^{(a)}$}
\address{$^{(a)}$ Facultad de Ciencias F\'{\i}sico-Matem\'aticas, Benem\'erita Universidad Aut\'onoma de Puebla,
Apartado Postal 1364, C.P. 72000 Puebla, Pue., M\'exico.\\
$^{(b)}$ Facultad de Ciencias de la Electr\'onica, Benem\'erita
Universidad Aut\'onoma de Puebla,
 Apartado Postal 1152, C.P. 72570 Puebla, Pue., M\'exico.\\
$^{(c)}$ Instituto de F\'{\i}sica, Benem\'erita Universidad
Aut\'onoma de Puebla, Apartado Postal J-48, C.P. 72570 Puebla, Pue.,
M\'exico.}
\date{\today}
\begin{abstract}
We explore the phenomenology of the Higgs sector in a model
that includes right-handed neutrinos, with a mass of the order of the
electroweak scale. In this model all scales arise from spontaneous
symmetry breaking, thus the Higgs sector
includes an extra Higgs singlet, in addition to the standard model
Higgs doublet. The scalar spectrum includes two neutral CP-even
states ($h$ and $H$, with $m_{h} < m_{H}$) and a neutral CP-odd
state ($\sigma$) that can be identified as a pseudo-Majoron. The
parameter of the Higgs potential are constrained using a
perturbativity criteria, which amounts to solve the corresponding RGE. The relevant
Higgs BR's and some cross-sections are discussed, with special
emphasis on the detection of the invisible Higgs signal at the LHC.
\end{abstract}
\maketitle

\setcounter{footnote}{0}
\setcounter{page}{2}
\setcounter{section}{0}
\setcounter{subsection}{0}
\setcounter{subsubsection}{0}

\section{Introduction.}

Neutrino physics has received a tremendous amount of experimental
input in the last
decade~\cite{neutrinoresults1,neutrinoresults2,neutrinoresults3,
neutrinoresults4,neutrinoresults5,neutrinoresults6,neutrinoresults7}.
Neutrino oscillations could be considered the first signal of
physics beyond the Standard Model (SM) \cite{stanmod}, and they
suggest that neutrinos are massive. After the data on
atmospheric and accelerator neutrino oscillations, we know that
there is a non-vanishing mass difference \cite{fukuda}. From solar and reactor
neutrino oscillations, we know that at least two neutrinos are not
massless \cite{neuoscillation}. On the theoretical side, the origin
of neutrino masses and their observed patterns (for the neutrino
mass squared differences) as well as the mixing angles still
represent a mystery~\cite{vallerev}. There are some ideas that have
been widely used in order to explore the situation, like the Zee
model~\cite{zee} or the see-saw mechanism~\cite{seesaw1,seesaw2} in
its several incarnations~\cite{seesaw3}, but we are far from a deep
understanding of this issue. Most of the actual realizations of
these mechanisms postpone the desired knowledge up to very
high, experimentally unaccessible, energy scales. Concretely, since
the introduction of Right-handed (RH) neutrinos seem to be the
obvious addition needed in order to write a Dirac mass for the
neutrinos, most models assume their existence with a mass scale
typically of size $\sim 10^{13}-10^{16}$~GeV (and the seesaw
mechanism can be used to explain the smallness of the neutrino mass
scale) ~\cite{seesaw2,seesaw3}.

In this paper we adhere to the idea that our current (experimental)
knowledge of particle physics should be explored by a "truly
minimal" extension of the SM. In this tenor we consider the
possibility of having only one scale associated with all the high
energy physics (HEP) phenomena. Since the SM is consistent with all
data so far (modulo neutrino masses), we propose a minimal extension
of the SM where new phenomena associated to neutrino physics can
also be explained by physics at the Electroweak (EW) scale {\it i.e.}
$O(10)$~GeV to $O(1)$~TeV (similar approaches
can be found in~\cite{Aranda:2007dq,similar1,similar2,similar3}).
Thus, we assume
\begin{itemize}
    \item SM particle content and gauge interactions.
    \item Existence of three RH neutrinos with a mass scale of
    EW size.
    \item Global U(1)$_L$ spontaneously (and/or explicitly) broken
    at the EW scale by a single complex scalar field.
    \item All mass scales come from spontaneous symmetry breaking
    (SSB). This leads to a Higgs sector that includes a Higgs
    SU(2)$_L$ doublet field $\Phi$ with hypercharge $1$
    (i.e. the usual SM Higgs doublet) and a SM singlet complex
    scalar field $\eta$ with lepton number $-2$.
\end{itemize}

This approach will have an effect on the type of signals usually
expected for the SM Higgs sector, where the hierarchy (naturalness)
problem resides. By enlarging the SM to explain the neutrino
experimental results, we can get a richer spectrum of signals for
Higgs physics and it is expected that once the LHC starts, it will
test some of the theoretical frameworks created thus
far, including ours. Furthermore, in order to fully probe whether
the Higgs bosons have ``Dirac'' and/or ``Majorana'' couplings, we
might have to wait until we reach a ``precision Higgs era'' at a
linear collider~\cite{lc}.

In this work, we explore in detail the Higgs phenomenology that
results in this model, with right-handed neutrinos having a mass
scale of the order of the electroweak scale. The scalar spectrum
includes two neutral CP-even states ($h$ and $H$ with $m_{h} <
m_{H}$) and a neutral CP-odd state ($\sigma$) that can be identified
as a pseudo-Majoron. The parameter of the Higgs potential are
constrained using a perturbativity criteria, which requires solving
the corresponding RGE. We then evaluate the dominant BR's and some
cross sections, with special emphasis on the detection of the
invisible Higgs signal at the LHC.

Our paper is organized as follows. In section II, we discuss the Lagrangian of our
model. In section III, we present the constraints on the model
parameters obtained by using the Renormalization Group Equations. In
section IV, we give the formulae to calculate the Higgs
decays, while in section V, we discuss the possibility of detecting
the invisible Higgs signal at the LHC. Finally, in section VI, we
summarize our results and present some conclusions.

\section{The model}

Taking into account the previous assumptions it is straightforward
to write the Lagrangian of the model. The relevant terms for Higgs
and neutrino sectors are
\beq \label{lagrangian}
    {\cal L}_{\nu H}={\cal L}_{\nu y} - V \ ,
\eeq
with
\beq \label{yukawas}
    {\cal L}_{\nu y}= -y_{\alpha i}\bar{L}_{\alpha}N_{Ri}\Phi
    -\frac{1}{2}Z_{ij}\eta \bar{N}_{Ri}^cN_{Rj} + h.c. \ ,
\eeq
where $N_R$ represents the RH neutrinos, $\psi^c=C\gamma^0\psi^*$
and $\psi_R^c\equiv(\psi_R)^c=P_L\psi^c$ has left-handed chirality.
The Yukawa couplings will be adjusted to reproduce the neutrino masses.
The potential of the Higgs sector is given by
\begin{eqnarray}
\label{potential}\nonumber
    V & = &
    \mu_D^2\Phi^{\dagger}\Phi+\frac{\lambda}{2}\left(\Phi^{\dagger}\Phi\right)^2
    + \mu_S^2\eta^*\eta +
    \lambda^{\prime}\left(\eta^*\eta\right)^2  \\
    & + & \kappa \left(\eta\Phi^{\dagger}\Phi + h.c. \right)
    +\lambda_m \left(\Phi^{\dagger}\Phi\right)\left(\eta^*\eta\right) \ .
\end{eqnarray}
Note that the fifth term in the potential (proportional to $\kappa$) breaks explicitly the
U(1) symmetry associated to lepton number. This is going to be relevant when
we discuss the features of the Majoron later in the paper.

Assuming that the scalar fields acquire vacuum expectation values
(vevs) in such a way that $\Phi$ and $\eta$ are responsible for EW
and global U(1)$_L$ symmetry breaking, respectively, we can write the
shifted fields (in unitary gauge)
\beq \label{vevs}
    \Phi = \left( \begin{array}{c} 0 \\ \frac{\phi^0+ v}{\sqrt{2}}
    \end{array} \right) \ \ {\rm and} \ \ \eta = \frac{\rho + u + i
    \sigma}{\sqrt{2}} \ ,
\eeq where $v/\sqrt{2}$ and $u/\sqrt{2}$ are the vevs of $\Phi$ and
$\eta$, respectively. Then, we obtain the following minimization
conditions: \begin{eqnarray}\label{minimization}
    \mu_D^2 & = & -\frac{1}{2} \left( \lambda v^2 + \lambda_m u^2
    -2\sqrt{2} \kappa u \right) \\
    \mu_S^2 & = & -\frac{1}{2u} \left( 2\lambda^{\prime} u^3 +
    \lambda_m u v^2 + \sqrt{2} \kappa v^2 \right) \ .
\end{eqnarray}
The form of the mass matrix for the scalar fields is given by
\beq \label{scalarmass}
    m_S^2= \left(\begin{array}{cc}
    \lambda v^2 & v u(\lambda_m  - \sqrt{2} r) \\ v u(\lambda_m
    -\sqrt{2} r) & 2\lambda^{\prime}u^2+\frac{1
    }{\sqrt{2}}r v^2 \end{array}\right) \ ,
\eeq where $r\equiv -\kappa/u$. The mass for the $\sigma$
(pseudo-Majoron) field is \beq \label{sigmamass}
m_{\sigma}^2=\frac{r v^2}{\sqrt{2}} \ . \eeq Note that, as expected,
$m_{\sigma}^2$ is proportional to the parameter $\kappa$ associated
to the explicit breaking of the U(1)$_L$ symmetry.

We are working under the assumption that the explicit breaking is
quite small, i.e. $\kappa <<$ EW scale. This explains why we are
minimizing the potential with respect to $\eta$, thus assuming it
breaks the global symmetry spontaneously. Furthermore we expect the
SSB generated by the vev of $\langle \eta \rangle = u/\sqrt{2}$ to
be of EW scale size, and so we work under the assumption $r\equiv
-\kappa/u << 1$. For example, taking $-\kappa \sim$~KeV one obtains
$r \sim 10^{-9}-10^{-7}$, which then leads to a Majoron mass of
$O(10)$ MeV.

From Eq.(\ref{scalarmass}), we can obtain the mass eigenstates
\beq \label{scalarfield}
     {\cal H} = \left( \begin{array}{c}
     \phi^0 \\ \rho \end{array} \right) =
     \left( \begin{array}{cc}
     \cos\alpha & -\sin\alpha \\ \sin\alpha & \cos\alpha \end{array} \right)
     \left( \begin{array}{c}
     h \\ H \end{array} \right) \ .
\eeq

Using these definitions to rewrite the Yukawa Lagrangian (Eq.(\ref{yukawas})), we obtain
\begin{eqnarray}
\label{yukawas2} \nonumber
     {\cal L}_{\nu y} & \supset & -y_{\alpha i} \bar{\nu}_{L\alpha}N_{Ri}\frac{\phi^0}{\sqrt{2}}
     -\frac{1}{2} Z_{ij} \frac{(\rho+i\sigma)}{\sqrt{2}} \bar{N}_{Ri}^c N_{Rj} + h.c. \\ \nonumber
     & = & \left( -\frac{y_{\alpha i}}{\sqrt{2}}\bar{\nu}_{L\alpha}N_{Ri}
     (c_{\alpha} \ h -s_{\alpha} \ H) + h.c. \right) - \left(\frac{i}{2\sqrt{2}}
     Z_{ij} \bar{N}_{Ri}^c N_{Rj} \sigma + h.c. \right) \\
     & - & \left( \frac{1}{2\sqrt{2}} Z_{ij} \bar{N}_{Ri}^c N_{Rj} (s_{\alpha} \ h +
     c_{\alpha} \ H) + h.c \right) \ .
\end{eqnarray}

We now make some comments regarding neutrino mass scales. Since we
are interested in RH neutrinos at the EW scale, we take their masses
to be in that scale, i.e. anywhere from a few to hundreds of GeV.
The Dirac part on the other hand will be constrained from the implementation
of the seesaw mechanism. The neutrino mass matrix is given by
\beq \label{mneutrino} m_{\nu} = \left( \begin{array}{cc} 0 & m_D \\
m_D & m_M \end{array}\right) \ , \eeq where $(m_D)_{\alpha i} =
y_{\alpha i}v/\sqrt{2}$. This is a $6 \times 6$ matrix, difficult to
analyze in general. But as an example, let us consider the third
family of SM fields and one RH neutrino, thus Eq.({\ref{mneutrino})
becomes a $2\times 2$ matrix. Assuming $m_D << m_M$ we obtain the
eigenvalues $m_1=-m_D^2/m_M$ and $m_2=m_M$; then by requiring $m_1
\sim$~O(eV), $m_2 \sim (10 -100)$~GeV, and using $v=246$~GeV, we
obtain an upper bound estimate for the Yukawa coupling $y_{\tau i} \leq
10^{-6}$.

The neutrino mass eigenstates are denoted by $\nu_1$ and $\nu_2$ and are defined such that
\begin{eqnarray} \label{transformation} \nonumber
\nu_{\tau} & = & \cos\theta \ \nu_{L1} + \sin\theta \ \nu_{R2} \\
N & = & -\sin\theta \ \nu_{L1} +\cos\theta \ \nu_{R2} \ ,
\end{eqnarray} where $\theta=\sqrt{m_D/m_2} \sim 10^{-6}-10^{-5}$.

The relevant terms in the Lagrangian become \begin{eqnarray}
\label{relevantterms} \nonumber {\cal L} & \supset & \left[ h
\bar{\nu}_{L1}^c \nu_{L1} \left( -\frac{Z}{2\sqrt{2}}s_\theta^2
s_\alpha \right)+  h \bar{\nu}_{R2}^c \nu_{R2}
\left(- \frac{Z}{2\sqrt{2}} c_\theta^2 s_\alpha \right) + h.c. \right] \\
& + & h \bar{\nu}_{L1} \nu_{R2}
\left(\frac{y_\nu}{\sqrt{2}}(s_{\theta}^2-c_{\theta}^2)c_\alpha
\right) +  h \bar{\nu}_{R2} \nu_{L1}
\left(\frac{y_\nu}{\sqrt{2}}(s_\theta^2-c_{\theta}^2) c_\alpha
\right) \ , \end{eqnarray} where $y_{\nu}^*=y_{\nu}$ and $Z \equiv Z_{11}$.

As discussed in the introduction we are also interested in exploring
the Higgs decay mode involving the Majoron. Then, we need to rewrite
to the terms in the potential that contain the Majoron-Higgs
bosons couplings, in terms of mass eigenstates. We obtain:

\begin{eqnarray}
V &\supset& \frac{1}{2} (c_{\alpha} v \lambda_{m} + 2 s_{\alpha} u
\lambda') h\sigma \sigma - \frac{1}{2} (s_{\alpha} v \lambda_{m} - 2 c_{\alpha} u \lambda') H\sigma \sigma \nonumber \\
&+& \frac{1}{4} (c_{\alpha}^{2} \lambda_{m} + 2 s_{\alpha}^{2}
\lambda')hh\sigma \sigma - \frac{1}{2}
c_{\alpha}s_{\alpha}(\lambda_{m} - 2 \lambda')
Hh\sigma \sigma \nonumber \\
&+& \frac{1}{4} (s_{\alpha}^{2} \lambda_{m} + 2 c_{\alpha}^{2}
\lambda')HH\sigma \sigma
\end{eqnarray}

\section{Constraints on the model parameters using RGE}

The parameters that appear in the Higgs potential are essentially
unconstrained by any phenomenology argument, therefore we have to
resort to some theoretical argument. Here we shall rely on the
perturbativity criteria, namely we shall require that any choice of
the Higgs parameters ($\lambda_i$) at low-energies, which determine the Higgs spectrum
and couplings, must remain below $4 \pi$ when evolve from $m_Z$ up
to a high-energy scale (such as $M_{GUT}$ or $M_{Planck}$).

The corresponding Renormalization Group Equations have been discussed in a slightly different context in
\cite{Basso:2010jm,Bassotesis}. They are given by

\begin{eqnarray}
\frac{dg_s}{dt}&=& \frac{g^3_s}{16 \pi^2} [-11 + \frac{4}{3}n_g] \nonumber \\
\frac{dg}{dt}&=& \frac{g^3}{16 \pi^2} [-\frac{22}{3} + \frac{4}{3}n_g +\frac{1}{6}] \nonumber \\
\frac{dg'}{dt}&=& \frac{g'^3}{16 \pi^2} [A^{YY}] \nonumber \\
\frac{dy_t}{dt}&=& \frac{y_t}{16 \pi^2} [\frac{9}{2} y^2_t - 8 g^2_s -\frac{9}{4} g^2 -\frac{17}{2} g'^2] \nonumber \\
\frac{dy^M_i}{dt}&=& \frac{y^M_i}{16 \pi^2} (4 (y^M_i)^2 + 2
Tr[(y^M)^2]), \,\,\,\,\, (i=1,2,3) \nonumber \\
\frac{d\lambda_1}{dt}&=& \frac{1}{16 \pi^2} (24 \lambda^2_1 +
\lambda^2_3 - 6 y^4_t +\frac{9}{8} g^4+ \frac{3}{8} g'^4 +
\frac{3}{4} g^2 g'^2 + 12 \lambda_1 y^2_t - 9 \lambda_1 g^2 -3 \lambda_1 g'^2) \nonumber \\
\frac{d\lambda_2}{dt}&=& \frac{1}{16 \pi^2} (20 \lambda^2_2 + 2
\lambda^2_3 - Tr[(y^M)^4] + 8 \lambda_2 Tr[(y^M)^2] \nonumber \\
\frac{d\lambda_3}{dt}&=& \frac{\lambda_3}{16 \pi^2} (12 \lambda_1 +
8 \lambda_2 + 4 \lambda_3 + 6 y^2_t - \frac{9}{2} g^2 - \frac{3}{2}
g'^2 + 4 Tr[(y^M)^2]) \nonumber
\end{eqnarray}

\noindent where $n_g$ stands by the family number (we are taking
$n_g=3$) and $y^M \equiv diag(y^M_1,y^M_2,y^M_3)$.

We have explored the values of parameters $\lambda$, $\lambda '$, $\lambda_m$, which satisfy the
perturbative constraint $g_i (M_{GUT})$, $\lambda_i (M_{GUT})$ $< 4 \pi$, and we identify some
scenarios that are safe to be studied in the following sections. Namely, we identify the following examples:

\begin{enumerate}
\item For $g_{top}(100 \gev) \approx 1$, and $y^M_3(100 \gev)=\lambda(100
\gev)=\lambda'(100 \gev)=\lambda_m(100 \gev)=0.3$, (Set 1), we can see from Fig.~\ref{fig:running1}
that their evolution from the EW scale ($100$ GeV) remains perturbative.

\item On the other hand, for another set of parameters, $g_{top}(100 \gev) \approx 1$, and $y^M_3(100 \gev)=\lambda(100
\gev)=\lambda'(100 \gev)=\lambda_m(100 \gev)=0.4$ (Set 2), we find
that the couplings blow up at an scale of O($10^{12}$), as it is
shown in Fig.~\ref{fig:running2}. Thus, just by going from $0.3$ to
$0.4$, for the values of the parameters, we find a change of regime.
\end{enumerate}

In what follows, we shall use Set 1 for the parameters of the Higgs
potential. This scenario can be taken as an example where parameters have
the maximal values allowed by the perturbativity criteria. Lower values are
thus allowed too.

\begin{figure}[floatfix]
\begin{center}
\includegraphics{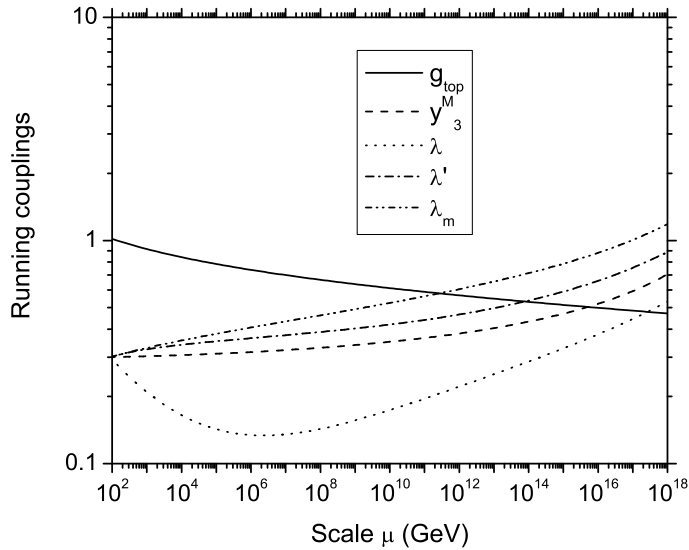}
\caption{Evolution of the running couplings: $g_{top}$, $y^M_3$,
$\lambda$, $\lambda'$ and $\lambda_m$ as functions of the scale
$\mu$, by taking $g_{top}(100 \gev) \approx 1$, and $y^M_3(100 \gev)=\lambda(100
\gev)=\lambda'(100 \gev)=\lambda_m(100 \gev)=0.3$
.}\label{fig:running1}
\end{center}
\end{figure}

\begin{figure}[floatfix]
\begin{center}
\includegraphics{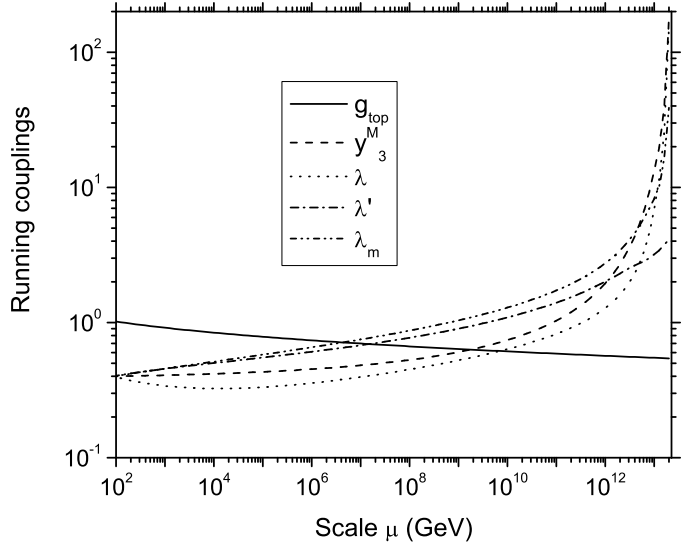}
\caption{Evolution of the running couplings: $g_{top}$, $y^M_3$,
$\lambda$, $\lambda'$ and $\lambda_m$ as functions of the scale
$\mu$, by taking $g_{top}(100 \gev) \approx 1$, and $y^M_3(100 \gev)=\lambda(100
\gev)=\lambda'(100 \gev)=\lambda_m(100 \gev)=0.4$
.}\label{fig:running2}
\end{center}
\end{figure}

\section{Higgs decays}

We are interested now in studying the new Higgs modes that appear in this model.
We have evaluated the Higgs decays using the formulae presented in
\cite{HHG}, with appropriate modifications to include the changes in
couplings due to Higgs mixing. The decay width for the Higgs decay mode into
a pair of majorons is given as follows:

\begin{equation}
\label{phidecay1} \Gamma (h \to \sigma \sigma)=
\frac{S|c_{h\sigma \sigma}|^{2} \sqrt{m_{h}^{4} - 4
m_{h}^{2} m_{\sigma}^{2}}}{16\pi m_{h}^{3}} ,
\end{equation}

\noindent where $c_{h\sigma\sigma}$ stands for the coupling $h
\sigma \sigma$; which was studied in the Section II. Namely $c_{h\sigma\sigma}
= - \frac{i}{2} (c_{\alpha} v \lambda_{m} + 2 s_{\alpha} u
\lambda')$, $c_{H\sigma\sigma} = \frac{i}{2} (s_{\alpha} v
\lambda_{m} - 2 c_{\alpha} u \lambda')$. Furthermore, in this case
$S=\frac{1}{2!}$.

In Ref.~\cite{Aranda:2007dq} we discussed the Higgs decays to
neutrinos and their signatures in this model. The possible relation to Majoron Dark Matter has been
considered in Ref.~\cite{Aranda:2009yb}.

Here, using Eq.~(\ref{relevantterms}) one obtains the following decay widths
\footnote{All SM decay widths will include an extra factor of
$c_{\alpha}^2$}: \begin{eqnarray} \label{widths}
\Gamma(h\to \bar{\nu}_1\nu_1) & = & \frac{m_h}{64\pi}|Z|^2 s_{\theta}^4 s_{\alpha}^2 \ , \\
\Gamma(h\to \bar{\nu}_2\nu_2) & = & \frac{m_h}{64\pi}|Z|^2
c_{\theta}^4 s_{\alpha}^2
\left(1-\frac{4m_2^2}{m_h^2}\right)^{3/2} \ , \\
\Gamma(h\to \bar{\nu}_1\nu_2) & = &
\frac{m_h}{16\pi}y_{\nu}^2(s_{\theta}^2-c_{\theta}^2)^2 c_{\alpha}^2
\left(1-\frac{m_2^2}{m_h^2}\right)^2 \ . \end{eqnarray}

In order to perform our numerical analysis, we shall take values
allowed by the perturbative analysis (Set 1) for
the parameters $\lambda_{m}$, $\lambda'$, and some
values for the vev $u$ and $\cos\alpha$. The mass of the Majoron
$m_{\sigma}$ is given in terms of the parameter $u$ as follows:
$$m_{\sigma}(u=10 \gev) \approx 65 \mev,$$
$$m_{\sigma}(u=246 \gev) \approx 13 \mev$$

For the light Higgs boson ($h$) we compute the BR's of the different
decay modes in the mass range $100 \leq m_{h} \leq 200$ GeV, assuming that the
heavier Higgs boson has a mass above 200 GeV, while the Majoron has a
mass of the order tenths of MeV.

\noindent In Figs. 3-6 we present our results for the decays: $h \to
b \bar{b}$, $h \to \tau \bar{\tau}$, $h \to W^+ W^-$, $h \to Z^0
Z^0$, $h \to \nu_2 \nu_2$, and $h \to \sigma \sigma$, which are the
most important two body decay modes. In Fig. 3 we show the
corresponding BR's considering $m_{\sigma}=13 \mev$ with $\lambda
'=\lambda_m=0.1$. Figs. 3(a), 3(b) and 3(c), corresponds to
$\cos\alpha=0.1, \, 0.5, \, \approx 1$, respectively. We can see
that $BR(h \to \nu_2 \nu_2)$ is dominant for $\cos
\alpha=0.1,\,0.5$, but it is of $O(10^{-7})$ for $\cos \alpha
\approx 1$. On the other hand, the $BR(h \to \sigma \sigma)$ is
relevant for any value of $\cos \alpha$, with a value of $O(10^{-2}
- 10^{-1})$ and becoming dominant when $\cos \alpha \approx 1$, in
the Higgs mass range $100 \lsim m_h \lsim 160 \gev$. We can see a
similar behavior in Fig. 4, where we are considering $m_{\sigma}=65
\mev$. The $BR(h \to \nu_2 \nu_2)$ is of the same order of magnitude
for a given value of $\cos\alpha$, and this quantity is independent
of the value of $m_{\sigma}$. On the other side, we observe in these
figures, that $BR(h \to \sigma \sigma)$ is sensitive to the value of
$m_\sigma$; when $m_\sigma=65 \mev$, this BR drops to
$O(10^{-4})$($O(10^{-3})$) for $\cos \alpha=0.1$($\cos \alpha=0.5$).
But it is still dominant for $\cos \alpha \approx 1$ in the mass
range $100 \lsim m_h \lsim 160 \gev$.

\noindent When we consider $\lambda '=\lambda_m=0.3$ (see Figs.
5-6), the $BR(h \to \nu_2 \nu_2)$ has the same behavior as in the
scenario with $\lambda '=\lambda_m=0.1$. However, in this case the
$BR(h \to \sigma \sigma)$ has an enhancement, becoming of
$O(10^{-1})$ for $\cos \alpha=0.1, 0.5$ and $m_\sigma=13\mev$, but
it is again dominant for $\cos \alpha \approx 1$, reaching values of
$O(10^{-1})$ for $100 \lsim m_h \lsim 160 \gev$. On the other hand,
the $BR(h \to \nu_2 \nu_2)$ in no longer dominant for $m_h \sim 100
\gev$ when $m_\sigma=13\mev$ and $\cos \alpha=0.1,\,0.5$. In the
case when $m_{\sigma}=65 \mev$, $BR(h \to \sigma \sigma)$ has
$O(10^{-3})$ ($O(10^{-2})$) for $\cos \alpha=0.1$ ($\cos
\alpha=0.5$) and, again, it is dominant in the case with $\cos
\alpha \approx 1$ for $100 \lsim m_h \lsim 160\gev$. If we compare
these BR's with the corresponding decay mode $h \to WW$. We can see
in Figs. 3-6 that the $BR(h \to WW)$ is sensitive to the value of
$\cos \alpha$, but not to the value of $m_\sigma$. It is also shown
in these figures that the decay mode to $WW$ has a BR of
$O(10^{-3})$ for $\cos \alpha=0.1$, and $O(10^{-2})$ for $\cos
\alpha=0.5$, and it becomes dominant when $\cos \alpha \approx 1$,
for any value of $m_{\sigma}$, and $160 \lsim m_h < 200 \gev$.
This behavior is realized for both $\lambda'=\lambda_m=0.1$ and $0.3$.

On the other hand, for the heavy Higgs boson ($H$) we consider the
mass range $150 \leq m_{H} \leq 500$, fixing the light Higgs
boson mass with two values $m_{h}=114$ GeV and $m_{h}=160$ GeV.

\noindent In Figs. 7-10 we present our results for the decays: $H
\to b \bar{b}$, $H \to \tau \bar{\tau}$, $H \to W^+ W^-$, $H \to Z^0
Z^0$, $H \to \nu_2 \nu_2$, and $H \to \sigma \sigma$, which are the most
important two body decay modes. In Fig. 7 we show the BR's considering $m_{\sigma}=13
\mev$ and $\lambda '=\lambda_m=0.1$. Figs. 7(a),
7(b) and 7(c), corresponds to $\cos\alpha=0.1, \, 0.5, \, \approx
1$, respectively. We can see that the $BR(h \to \nu_2 \nu_2)$ is
dominant for $\cos \alpha=0.1$, and it is no longer dominant when $m_H \gsim 250$
GeV for $\cos \alpha=0.5$, and it is of $O(10^{-7})$ for $\cos \alpha \approx 1$.
On the other hand, $BR(H \to \sigma \sigma)$ is relevant for any
value of $\cos \alpha$ and its relevance is larger when $\cos\alpha
\approx 1$. We can see a similar behavior for $BR(H \to \nu_2
\nu_2)$ and $BR(H \to \sigma \sigma)$ in Fig. 8, where we
have considered $m_{\sigma}=65 \mev$. The $BR(H \to \nu_2 \nu_2)$ is the dominant mode, when
$\cos\alpha=0.1$ independently of $m_{\sigma}$. The most important
decay mode is $BR(H \to \sigma \sigma)$, when $\cos\alpha \approx 1$
and it does not dependent on the $m_{\sigma}$ value.

\noindent When we consider $\lambda '=\lambda_m=0.3$ (see Figs. 9 and 10),
the $BR(h \to \sigma \sigma)$  shows an enhancement with
respect to the value shown in the previous case, where we are taking
$\lambda '=\lambda_m=0.1$. We can see in Figs. 9 and 10, a
similar behavior for $BR(H \to \nu_2 \nu_2)$ and $BR(H \to WW)$, when we consider
$m_{\sigma}=13 \mev$ and $m_{\sigma}=65 \mev$.
Namely, the $BR(H \to \nu_2 \nu_2)$ is the dominant
one, when $\cos\alpha=0.1$ independently of $m_{\sigma}$. The most
important decay mode is $BR(H \to \sigma \sigma)$, when $\cos\alpha
\approx 1$ and it does not dependent on the $m_{\sigma}$ value.
This behavior is realized for $\lambda '=\lambda_m=0.1, \, 0.3$

\section{Detection at LHC}

We are also interested in determining whether the invisible Higgs decay
could be observed at LHC. We shall use the results of
Ref.~\cite{Davoudiasl:2004aj}, where a detailed study of
detectability of an invisible Higgs was performed. These authors
consider the production mechanisms $pp \to h Z^0$, for the signal, as
well as $pp \to W^+ W^- \to h$ and $pp \to h + jet$. Here, we
shall use for illustration purposes, the
associated production with $Z^0 + h (\to inv.)$; a detailed set of
cuts is proposed in order to determine the backgrounds~\cite{Davoudiasl:2004aj}. Their analysis can be used to
determine the minimum value of the ratio

\begin{equation}
\label{rparameter}
R=\frac{g^2_{\phi ZZ}}{g^2_{\phi_{SM} ZZ}} \times BR (h \to \sigma \sigma)
\end{equation}

\noindent that can produce a $4 \sigma$ signal. This is shown in
Table~\ref{tab:t1} for two different values of the Luminosity ${\cal
L}$ and some values of $m_h$. We display in Table~\ref{tab:t2},
the value of $R$ for several choices of parameters within our model, and
it can be seen that these choices are consistent with the
perturbative analysis of the previous section. For instance, for $\lambda '=\lambda_m=0.3$,
and taking $\cos\alpha=0.99$, $u=246$ GeV and $m_H=160$, we obtain $R=0.846$. This
means that it coul be possible to find evidence of the existence of a
pseudo-Majoron through an invisible Higgs at the LHC for this scenario, with a significance of $4 \, \sigma$, even with
an integrated Luminosity of ${\cal L}=30 \, \ifb$.

\begin{table}
\begin{tabular}{|c|c|c|c|}
\hline  & \multicolumn{3}{c|}{Minimal value of $R$ to be observed at the LHC}\\
\cline{2-4}$\,$Luminosity$\;$ & $\, m_h=120 \gev$
& $\, m_h=140 \gev$ & $\, m_h=160 \gev$ \\
\hline \hline

$\; 30 \; \ifb \;$&$\;0.404 \; (0.439)\;$&$\;0.550 \; (0.598)\;$&$\;0.739 \; (0.803)\;$\\

$\; 50 \; \ifb \;$&$\;0.313 \; (0.340)\;$&$\;0.426 \; (0.463)\;$&$\;0.573 \; (0.622)\;$\\

$\; 100 \; \ifb \;$&$\;0.221 \; (0.241)\;$&$\;0.301 \; (0.328)\;$&$\;0.405 \; (0.440)\;$\\

\hline
\end{tabular}
\caption{Minimal value of the parameter $R$ for an event $h \to inv.$ to be observed with
a significance of $4 \, \sigma$ at the LHC, with an integrated Luminosity ${\cal L}$,
as a function of the Higgs boson mass $m_h$. Here we take the cut on $\slash{p}_T=75$ GeV of Ref. \cite{Davoudiasl:2004aj}. The
numbers in parentheses include the estimated $Z+jets$ background  discussed in Ref. \cite{Davoudiasl:2004aj}}\label{tab:t1}
\end{table}

\begin{table}
\begin{tabular}{|c|c|c|c|c|}
\hline & & \multicolumn{3}{c|}{$R=(g^2_{\phi
ZZ}/g^2_{\phi_{SM} ZZ}) \times BR (h \to inv.)$}\\
\cline{3-5} {$\, \cos \alpha$}& $u\;$ & $\, m_h=120 \gev$
& $\, m_h=140 \gev$ & $\, m_h=160 \gev$ \\
\hline \hline

$0.9$&$\;10 \gev\;$&$\;0.378\;$&$\;0.315\;$&$\;0.264\;$\\

$0.9$&$\;246 \gev\;$&$\;0.615\;$&$\;0.563\;$&$\;0.514\;$\\

$0.95$&$\;10 \gev\;$&$\;0.578\;$&$\;0.510\;$&$\;0.448\;$\\

$0.95$&$\;246 \gev\;$&$\;0.743\;$&$\;0.697\;$&$\;0.650\;$\\

$0.99$&$\;10 \gev\;$&$\;0.858\;$&$\;0.820\;$&$\;0.780\;$\\

$0.99$&$\;246 \gev\;$&$\;0.901\;$&$\;0.874\;$&$\;0.846\;$\\

\hline
\end{tabular}
\caption{Value of $R$ for the process $h \to \sigma \sigma$, by
fixing $\lambda' = \lambda_m = 0.3$ and  by taking several values of
the parameters $u= \sqrt{2} \, \langle \eta \rangle$ and
$\cos\alpha$, as a function of the Higgs boson mass
$m_h$.}\label{tab:t2}
\end{table}

\section{Conclusions}

We have explored in detail the Higgs phenomenology that results in a
model where right-handed neutrinos have a mass scale of the order of
the electroweak scale. In this model all scales arise from
spontaneous symmetry breaking, and this is achieved with a Higgs
sector that includes an extra Higgs singlet in addition to the
standard model Higgs doublet. The scalar spectrum includes two
neutral CP-even states ($h$ and $H$ with $m_{h} < m_{H}$) and a neutral
CP-odd state ($\sigma$) that can be identified as a
pseudo-Majoron.

The parameters that appear in the Higgs potential are essentially
unconstrained by any phenomenology argument, therefore we have to
resorted to some theoretical argument. Here we have relied on the
perturbativity criteria, namely we have required that any choice of
Higgs parameters at low-energies, which determine the Higgs spectrum
and couplings, must remain below $4 \pi$ when evolve from $m_Z$ up
to a high-energy scale (such as $M_{GUT}$ or $M_{Planck}$).

Thus, we have explored the values of parameters $\lambda$, $\lambda '$,
$\lambda_m$, which satisfy the perturbative constraint $g_i
(M_{GUT})$, $\lambda_i (M_{GUT})$ $< 4 \pi$, and we have identified
some safe scenarios that are studied in this paper.

We have concluded that Set 1 ($g_{top}(100 \gev) \approx 1$, and
$y^M_3(100 \gev)=\lambda(100 \gev)=\lambda'(100 \gev)=\lambda_m(100
\gev)=0.3$) can be taken as a scenario for the parameters of the
Higgs potential having the maximal values allowed by the perturbative analysis.

The relevant Higgs BR and cross-sections are discussed, with special
emphasis on the detection of the invisible Higgs signal at the LHC.
We conclude that it could be possible to detect evidence of the
existence of a pseudo-Majoron $\sigma$ through an invisible Higgs
signal at the LHC for some values of parameters.

\bigskip

\begin{center}
{\bf ACKNOWLEDGMENTS}
\end{center}

The authors are grateful to {\it Sistema Nacional de Investigadores
} and {\it CONACyT} (M\'{e}xico) for financial support.

\begin{figure}[floatfix]
\begin{center}
\includegraphics[height=12cm,width=15cm]{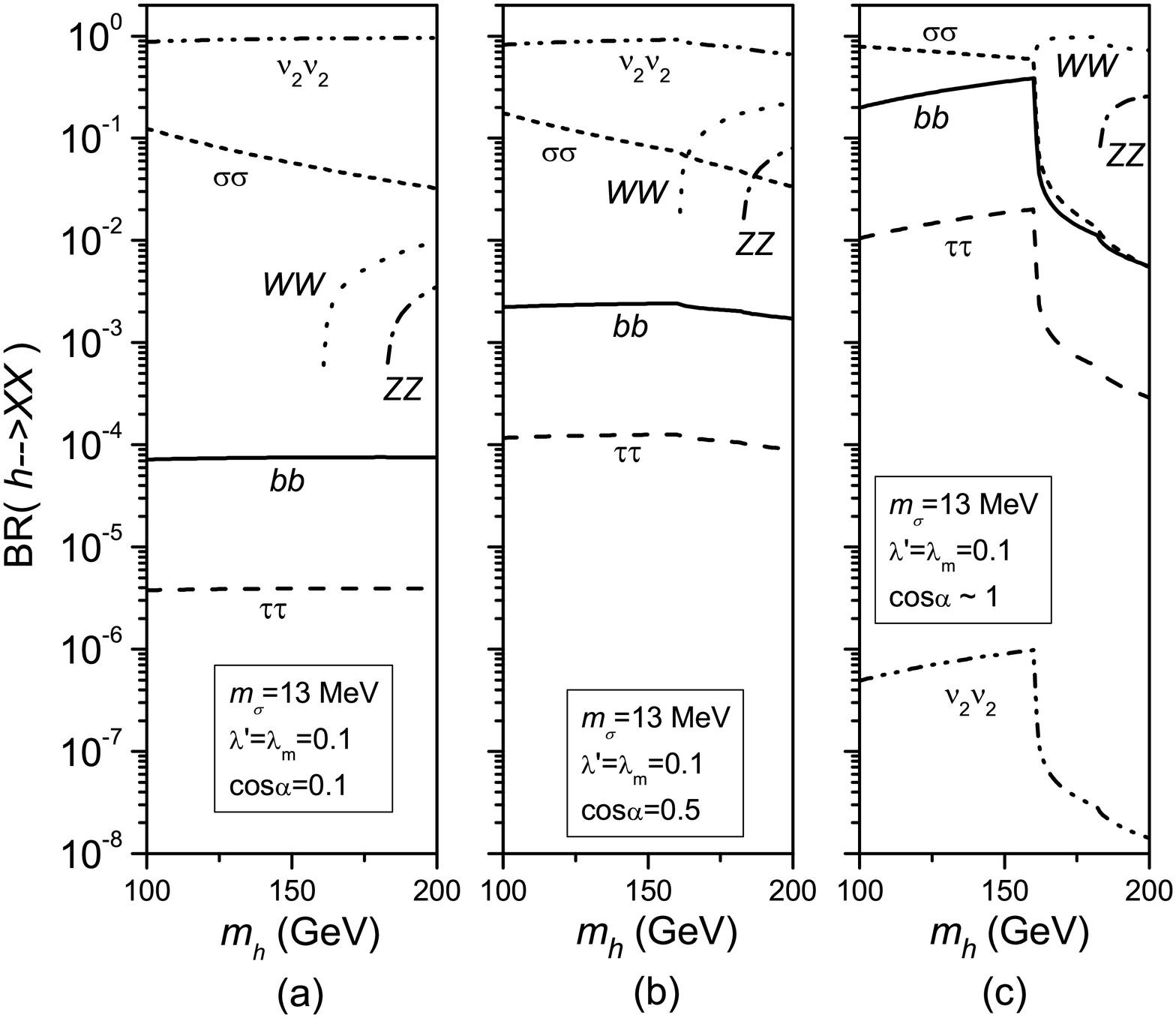}
\caption{Braching ratios for the decay $h\to XX$ with
$m_\sigma= 13 \mev$, $\lambda '=\lambda_m=0.1$ and three different
values for $\cos \alpha$: (a) $\cos \alpha=0.1$, (b) $\cos
\alpha=0.5$ and (c) $\cos \alpha \approx 1$.} \label{fig:h1}
\end{center}
\end{figure}

\begin{figure}[floatfix]
\begin{center}
\includegraphics[height=12cm,width=15cm]{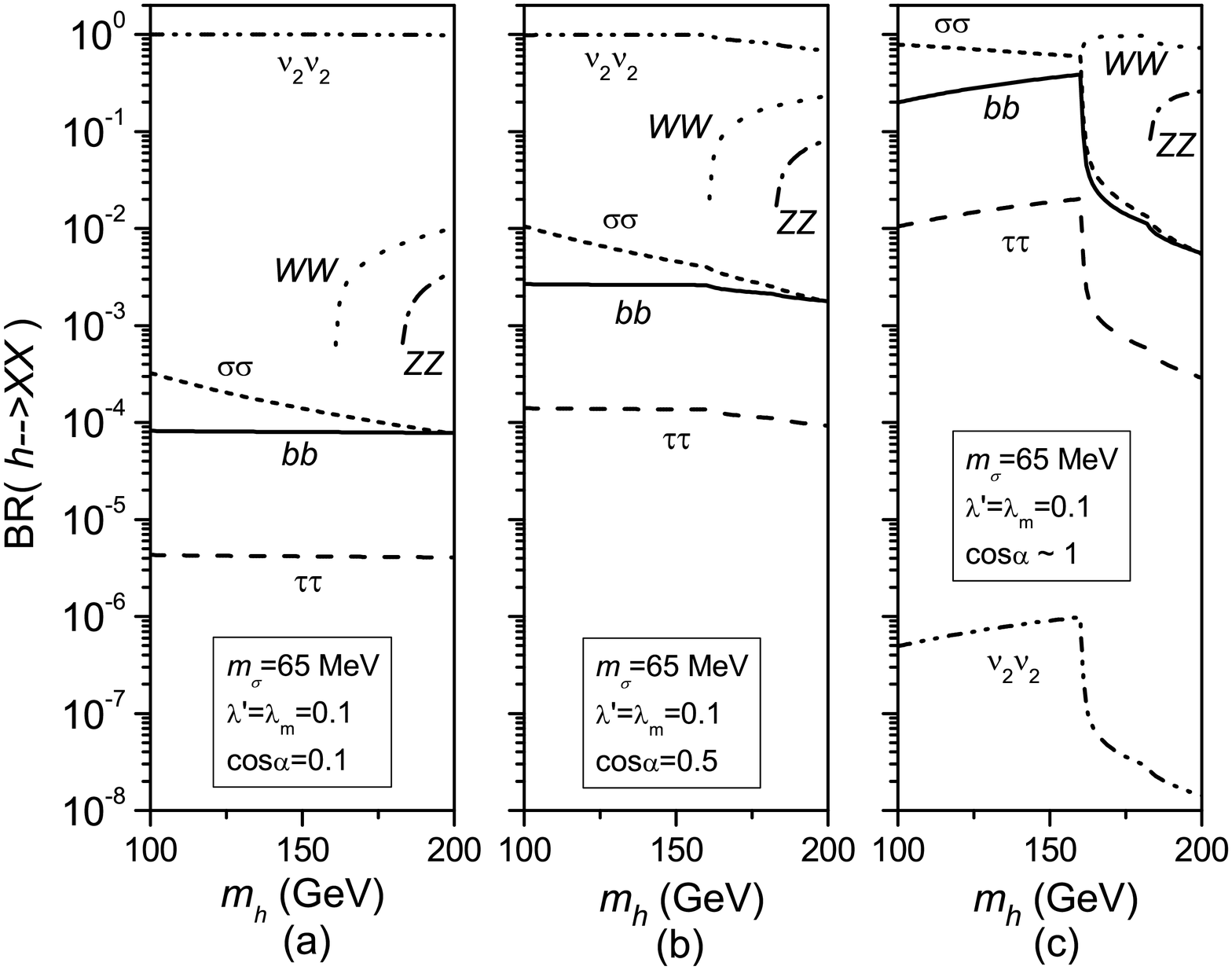}
\caption{Braching ratios for the decay $h\to XX$ with
$m_\sigma= 65 \mev$, $\lambda '=\lambda_m=0.1$ and three different
values for $\cos \alpha$: (a) $\cos \alpha=0.1$, (b) $\cos
\alpha=0.5$ and (c) $\cos \alpha \approx 1$.} \label{fig:h3}
\end{center}
\end{figure}

\begin{figure}[floatfix]
\begin{center}
\includegraphics[height=12cm,width=15cm]{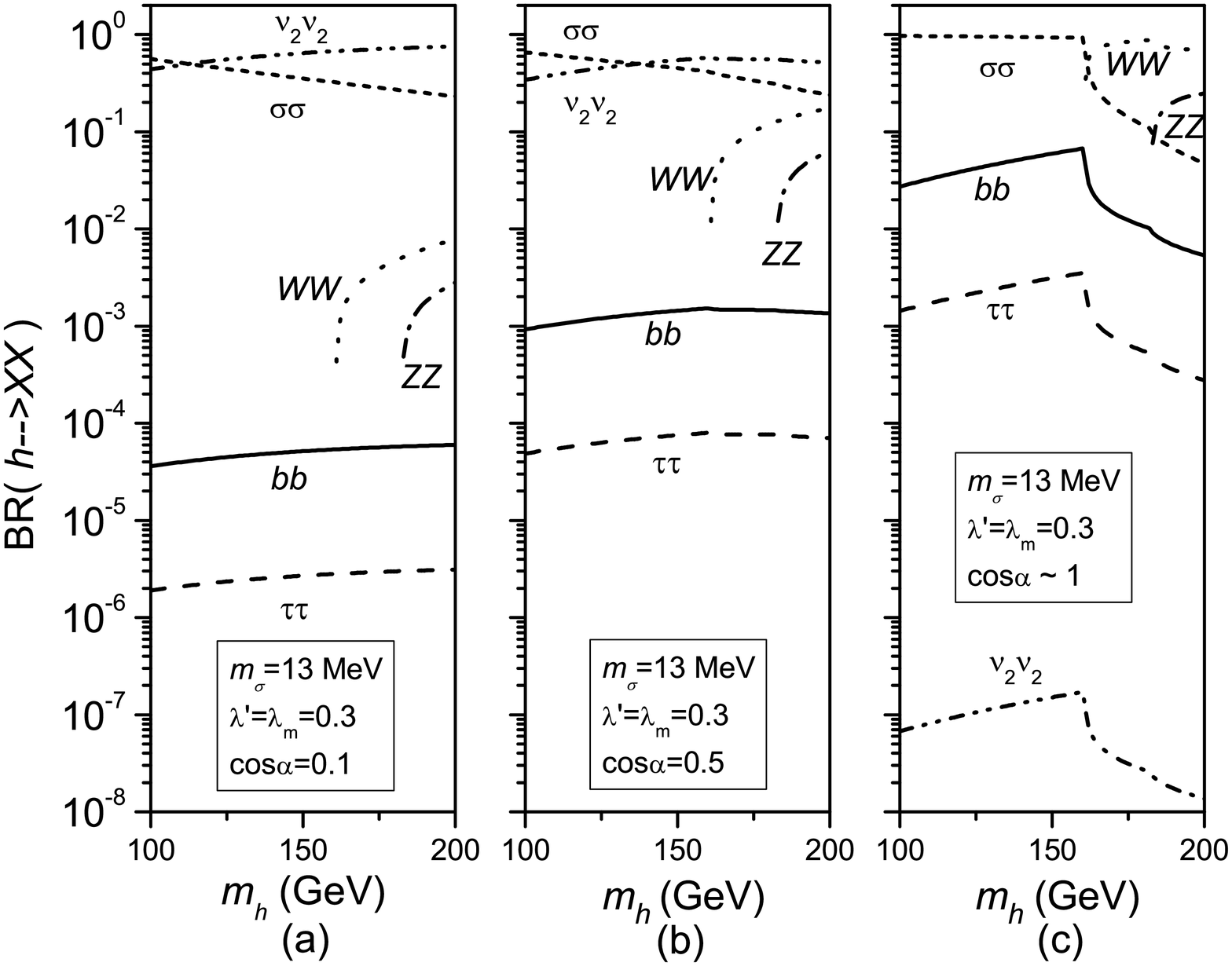}
\caption{Braching ratios for the decay $h\to XX$ with
$m_\sigma= 13 \mev$, $\lambda '=\lambda_m=0.3$ and three different
values for $\cos \alpha$: (a) $\cos \alpha=0.1$, (b) $\cos
\alpha=0.5$ and (c) $\cos \alpha \approx 1$.} \label{fig:h4}
\end{center}
\end{figure}

\begin{figure}[floatfix]
\begin{center}
\includegraphics[height=12cm,width=15cm]{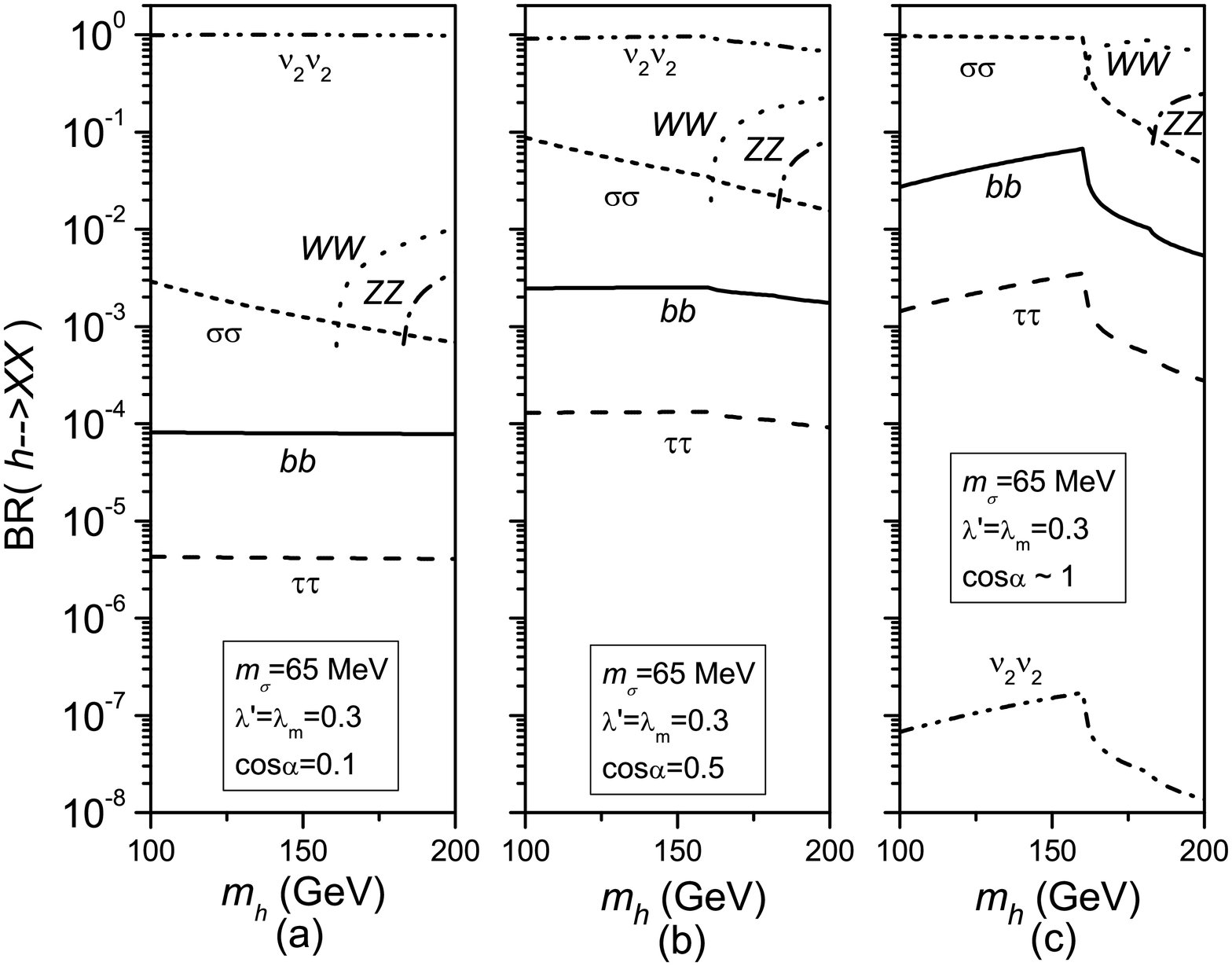}
\caption{Braching ratios for the decay $h\to XX$ with
$m_\sigma= 65 \mev$, $\lambda '=\lambda_m=0.3$ and three different
values for $\cos \alpha$: (a) $\cos \alpha=0.1$, (b) $\cos
\alpha=0.5$ and (c) $\cos \alpha \approx 1$.} \label{fig:h6}
\end{center}
\end{figure}

\begin{figure}[floatfix]
\begin{center}
\includegraphics[height=12cm,width=15cm]{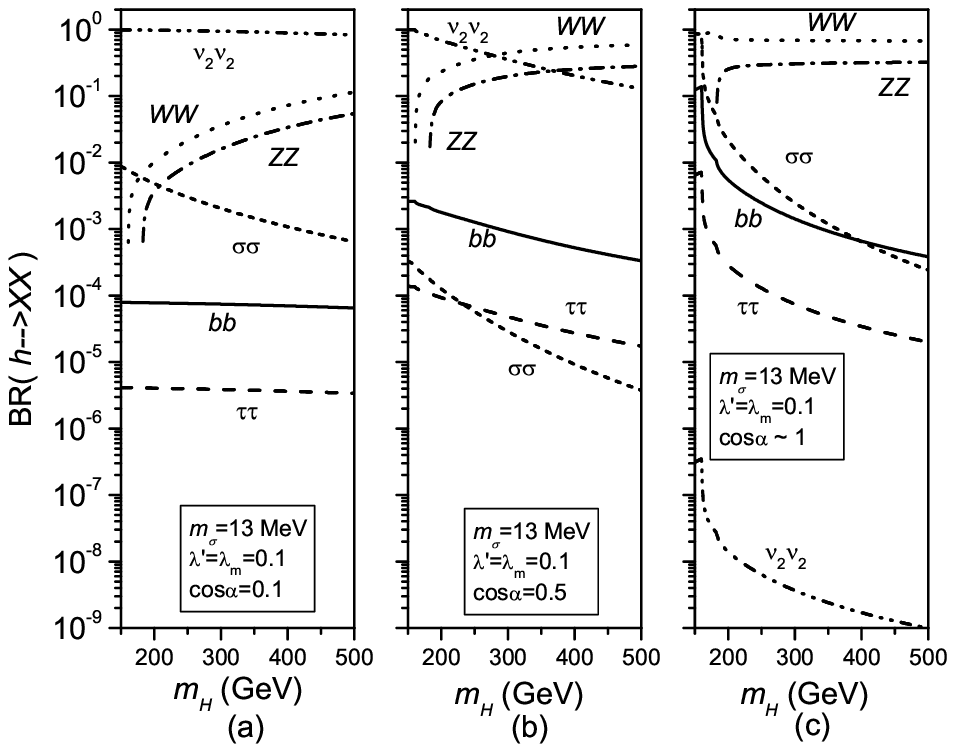}
\caption{Braching ratios for the decay $H\to XX$ with
$m_\sigma= 13 \mev$, $\lambda '=\lambda_m=0.1$ and three different
values for $\cos \alpha$: (a) $\cos \alpha=0.1$, (b) $\cos
\alpha=0.5$ and (c) $\cos \alpha \approx 1$.} \label{fig:hpes1}
\end{center}
\end{figure}

\begin{figure}[floatfix]
\begin{center}
\includegraphics[height=12cm,width=15cm]{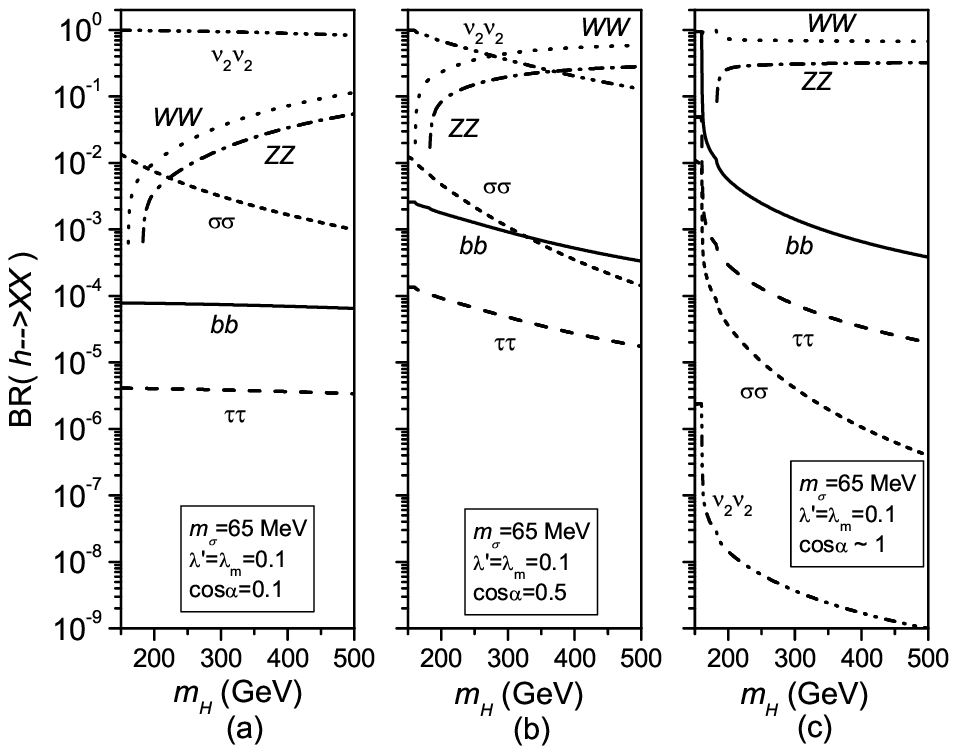}
\caption{Braching ratios for the decay $H\to XX$ with
$m_\sigma= 65 \mev$, $\lambda '=\lambda_m=0.1$ and three different
values for $\cos \alpha$: (a) $\cos \alpha=0.1$, (b) $\cos
\alpha=0.5$ and (c) $\cos \alpha \approx 1$.} \label{fig:hpes3}
\end{center}
\end{figure}

\begin{figure}[floatfix]
\begin{center}
\includegraphics[height=12cm,width=15cm]{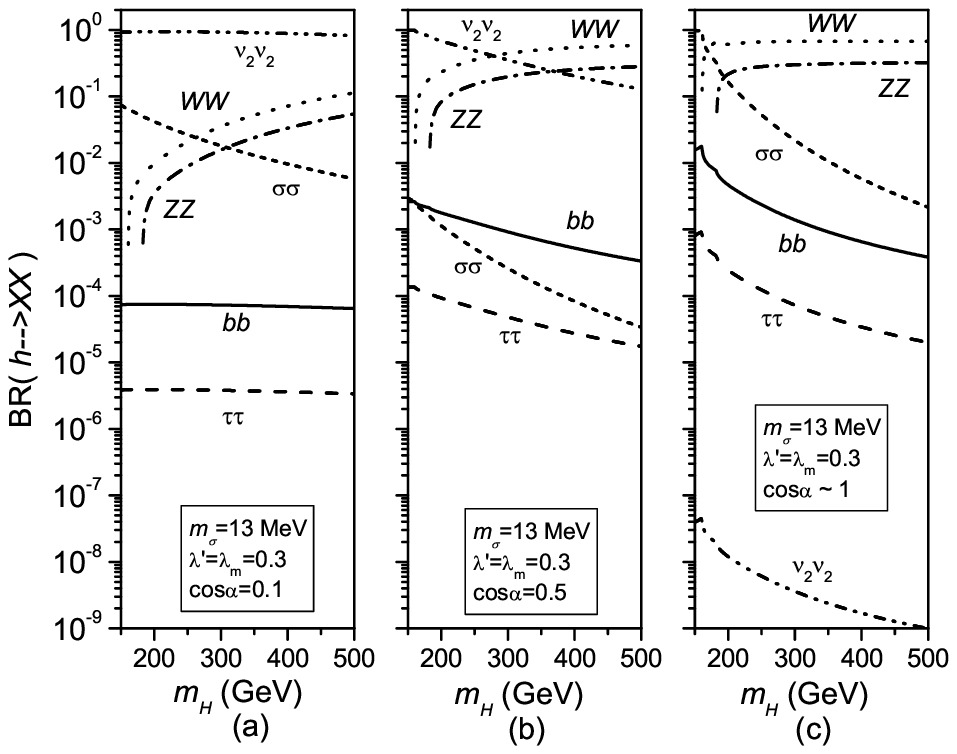}
\caption{Braching ratios for the decay $H\to XX$ with
$m_\sigma= 13 \mev$, $\lambda '=\lambda_m=0.3$ and three different
values for $\cos \alpha$: (a) $\cos \alpha=0.1$, (b) $\cos
\alpha=0.5$ and (c) $\cos \alpha \approx 1$.} \label{fig:hpes4}
\end{center}
\end{figure}

\begin{figure}[floatfix]
\begin{center}
\includegraphics[height=12cm,width=15cm]{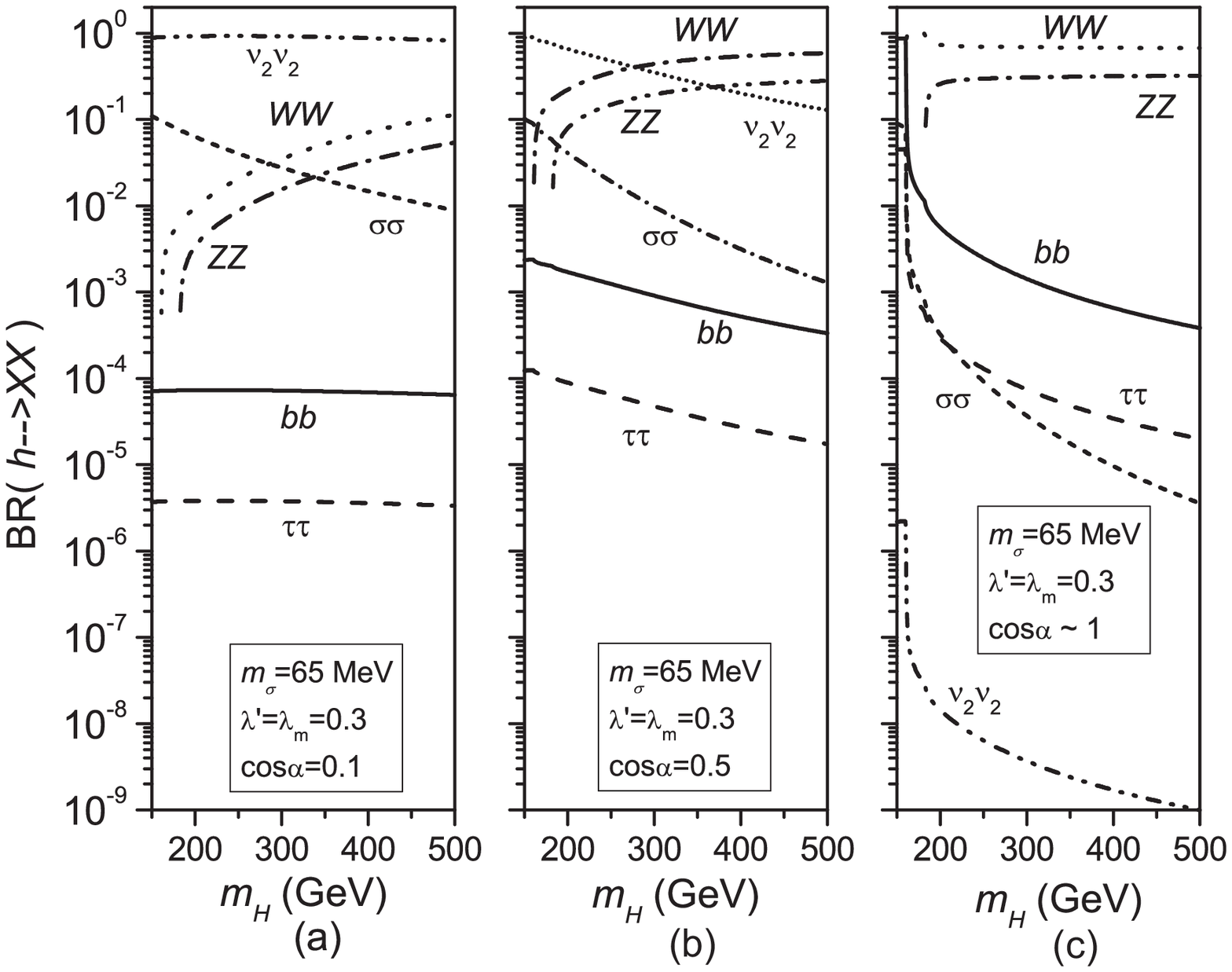}
\caption{Braching ratios for the decay $H\to XX$ with $m_\sigma= 65
\mev$, $\lambda '=\lambda_m=0.3$ and three different values for
$\cos \alpha$: (a) $\cos \alpha=0.1$, (b) $\cos \alpha=0.5$ and (c)
$\cos \alpha \approx 1$.} \label{hpes6}
\end{center}
\end{figure}

\end{document}